\documentclass[]{aa}
\usepackage[varg]{txfonts}

\usepackage[colorlinks=true,urlcolor=blue,citecolor=blue,linkcolor=blue]{hyperref}
\usepackage{twoopt}

\bibpunct{(}{)}{;}{a}{}{,} 

\makeatletter

 \newcommandtwoopt{\citeads}[3][][]{%
   \nonstopmode
   \href{http://adsabs.harvard.edu/abs/#3}%
        {\def\hyper@linkstart##1##2{}%
         \let\hyper@linkend\@empty\citealp[#1][#2]{#3}}
   \biblink{#3}{\href{http://adsabs.harvard.edu/abs/#3}{ADS}}%
   \errorstopmode}            
   
 \newcommandtwoopt{\citepads}[3][][]{%
   \nonstopmode
   \href{http://adsabs.harvard.edu/abs/#3}%
        {\def\hyper@linkstart##1##2{}%
         \let\hyper@linkend\@empty\citep[#1][#2]{#3}}
   \biblink{#3}{\href{http://adsabs.harvard.edu/abs/#3}{ADS}}
   \errorstopmode}            
   
 \newcommandtwoopt{\citetads}[3][][]{%
   \nonstopmode
   \href{http://adsabs.harvard.edu/abs/#3}
        {\def\hyper@linkstart##1##2{}%
         \let\hyper@linkend\@empty\citet[#1][#2]{#3}}
   \biblink{#3}{\href{http://adsabs.harvard.edu/abs/#3}{ADS}}%
   \errorstopmode}            
   
 \newcommandtwoopt{\citeyearads}[3][][]{%
   \nonstopmode
   \href{http://adsabs.harvard.edu/abs/#3}%
        {\def\hyper@linkstart##1##2{}%
         \let\hyper@linkend\@empty\citeyear[#1][#2]{#3}}
   \biblink{#3}{\href{http://adsabs.harvard.edu/abs/#3}{ADS}}%
   \errorstopmode}            
\makeatother

\makeatletter
\newcommand{\bibnote}[2]{\@namedef{#1note}{#2}}
\newcommand{\biblink}[2]{\@namedef{#1link}{#2}}
\makeatother

\def\Trad{\ensuremath{T_\mathrm{rad}}}
\def\CaII{\ion{Ca}{II}}
\def\CaIIK{\CaII~K}
\def\CaIIIR{\CaII~854.2~nm}
\def\CaIIHK{\CaII~H\&K}

\def\FeI{\ion{Fe}{I}}
\def\kone{{K$_1$}}
\def\ktwo{{K$_2$}}
\def\konev{K$_{1 \mathrm{v}}$}
\def\ktwor{K$_{2 \mathrm{r}}$}
\def\ktwov{K$_{2 \mathrm{v}}$}
\def\koner{K$_{1 \mathrm{r}}$}
\def\kthree{K$_{3}$}
\def\Halpha{\mbox{H$\alpha$}}

\newcommand{\dd}{\mathrm{d}}
\newcommand{\be}{\begin{equation}}
\newcommand{\ee}{\end{equation}}

\newcommand{\bea}{\begin{eqnarray}}
\newcommand{\eea}{\end{eqnarray}}
\newcommand{\edt}[1]{{#1}}

\begin{document}

\title{On chromospheric heating during flux emergence in the solar atmosphere}

\subtitle{ }

\author{Jorrit Leenaarts\inst{1} \and
 Jaime de la Cruz Rodr\'{i}guez\inst{1}  \and
 Sanja Danilovic\inst{1} \and
  G\"{o}ran Scharmer\inst{1}  \and
 Mats Carlsson\inst{2,3} }

\offprints{J. Leenaarts \email{jorrit.leenaarts@astro.su.se}}

\institute{
Institute for Solar Physics, Department of Astronomy, Stockholm University, AlbaNova University Centre, SE-106 91 Stockholm, Sweden
\and
Institute of Theoretical Astrophysics, University of Oslo, P.O. Box 1029 Blindern, N-0315 Oslo, Norway
\and
Rosseland Centre for Solar Physics, University of Oslo, P.O. Box 1029 Blindern, N-0315 Oslo, Norway}

\date{Received; Accepted }

\abstract 
{The radiative losses in the solar chromosphere vary from 4~kW~m$^{-2}$ in the quiet Sun, to 20~kW~m$^{-2}$ in active regions. The mechanisms that transport non-thermal energy to and deposit it in the chromosphere are still not understood.}
{We aim to investigate the atmospheric structure and heating of the solar chromosphere in an emerging flux region.}  
{We use observations taken with the CHROMIS and CRISP instruments on the Swedish 1-m Solar Telescope in the \CaIIK, \CaII~854.2~nm, H$\alpha$, and \FeI\ 630.1~nm and 630.2~nm lines. We analyse the various line profiles and in addition perform multi-line, multi-species, non-Local Thermodynamic Equilibrium (non-LTE) inversions to estimate the spatial and temporal variation of the chromospheric structure. }  
    {We investigate which spectral features of  \CaIIK\ contribute to the frequency-integrated \CaIIK\ brightness, which we use as a tracer of chromospheric radiative losses. The majority of the radiative losses are not associated with localized high-\CaIIK\-brightness events, but instead with a more gentle, spatially extended, and persistent heating. The frequency-integrated \CaIIK\ brightness correlates strongly with the total linear polarization in the \CaII~854.2~nm, while the \CaIIK\ profile shapes indicate that the bulk of the radiative losses occur in the lower chromosphere.
 Non-LTE inversions indicate a transition from heating concentrated around photospheric magnetic elements below $\log{\tau_{500}} =-3$ to a more space-filling and time-persistent heating above $\log{\tau_{500}} =-4$. The inferred gas temperature at $\log{\tau_{500}} =-3.8$ correlates strongly with the total linear polarization in the \CaIIIR\ line, suggesting that that the heating rate correlates with the strength of the horizontal magnetic field in the low chromosphere.  }
     {} 
     
    
    \maketitle


\section{Introduction} \label{sec:intro}

Active regions (ARs) are areas in the solar atmosphere with a high average magnetic flux density. They appear brighter than the surrounding atmosphere in most chromospheric and coronal diagnostics
\citepads[e.g.,][]{1975scco.conf.....A}.
The enhanced brightness is indicative of enhanced radiative energy losses. Typical estimates for the radiative energy losses in the quiet sun are 4~kW~m$^{-2}$, while for active regions they are 20~kW~m$^{-2}$
\citepads[e.g,][]{1977ARA&A..15..363W,1981ApJS...45..635V}.

The chromospheric radiative losses per volume in 1D models are largest just above the temperature minimum, implying that the largest energy deposition rate occurs there too. The temperature minimum itself has a temperature close to its radiative equilibrium value in models like VALC
\citepads{1974SoPh...39...19H,1981ApJS...45..635V,1992A&A...253..567C,1993A&A...275..101E}.
If the temperature there is larger, then even the temperature minimum is losing energy through radiative cooling.

Emerging flux regions harbor a variety of events, which in areas of limited extent in time and space lead to strong chromospheric heating, such as Ellerman Bombs (EBs) and UV bursts
\citepads[e.g.][]{2014Sci...346C.315P,2017ApJ...836...63T},
a picture that has been roughly confirmed by numerical simulations
\citepads{2017A&A...601A.122D,2017ApJ...839...22H}.

An important aspect of flux emergence in the chromosphere is the interaction of the emerging magnetic field with itself. Flux typically appears on the surface on granular scales in a sea-serpent configuration and it can get rid of its mass and emerge further through self-reconnection in the photosphere
\citepads{2010ApJ...720..233C}.
In the chromosphere it will interact with previously emerged loops that might have a different direction, and is then in a favorable configuration for Joule dissipation and/or reconnection to take place. MHD modelling of flux emergence and the resulting formation of current sheets upon interactions with pre-existing ambient magnetic field was investigated by
\citetads{2005ApJ...618L.153G}.
For a detailed overview of the observational and theoretical aspects of flux emergence see
\citetads{2014SSRv..186..227S}
and
\citetads{2014LRSP...11....3C}.
\citetads{2014ApJ...781..126O,2015ApJ...810..145D},
 and
\citetads{2016ApJ...825...93O}
discuss observations of the emergence and rise through the chromosphere of one small-scale loop in detail.

Heating in resistive MHD must happen through dissipation of currents 
Numerical simulations indicate that resistive MHD is not sufficient to accurately describe the partially ionized solar chromosphere.
Radiation-MHD models of emerging flux that employ standard resistive MHD typically give rise to  cold ($<2300$~K) bubbles in the upper photosphere and low chromosphere at heights of roughly 300\,--\,1000~km above $\tau_{500} =1$ 
\citepads{2008ApJ...679..871M,2009A&A...507..949T,2009ApJ...702..129M,2014ApJ...781..126O}.
The MHD codes must employ an artificial extra heating term to prevent the temperature in these bubbles from falling further. \edt{Cold bubbles are also observed, but inversions do not retrieve the same low temperatures as in the simulations
\citepads{2015ApJ...810..145D}.
}

%
%
%
%

A step beyond standard resistive MHD is the inclusion of some plasma physics processes in the models, especially collisions between ion and neutral atoms that create an anisotropic resistivity. The resulting resistivity for cross-field current (Pedersen resistivity) is much larger than the Spitzer resistivity and leads to strongly enhanced energy dissipation and increased  temperature in the chromosphere in simplified simulations of flux emergence
\citepads{2006A&A...450..805L,2007ApJ...666..541A,2013ApJ...764...54L}.

More realistic radiation-MHD simulations also show that inclusion of ion-neutral interaction effects leads to an increase in dissipation of magnetic energy and increased chromospheric temperatures
\citepads{2012ApJ...753..161M, 2012ApJ...747...87K,2015RSPTA.37340268M,2016ApJ...819L..11S,2017arXiv170806781M}.
However, such realistic radiation-MHD simulations including strong flux emergence and ion-neutral interactions have so far not been reported.

There are still many open questions associated with flux emergence through the chromosphere. The main question is still: how is the energy transported to and dissipated in the chromosphere to compensate for the large observed radiative losses. The magnetic field obviously plays an important role in these processes.

\citetads{1975ApJ...200..747S}
and
\citetads{1989ApJ...337..964S}
demonstrated correlation of \CaIIK\ brightness and photospheric vertical magnetic field. 
\edt{They used $\sim$110~pm wide filters centered on the line \CaIIK\ core, and the intensity that they employed thus contains considerable contamination from photospheric signal.}
The chromosphere of emerging flux regions contains much more horizontal  field than vertical field, but to our knowledge no studies have been performed studying the relation between the horizontal field and atmospheric structure and radiative losses. Here we report on an emerging flux region observed with the Swedish 1-m Solar Telescope 
\citepads[SST,][]{2003SPIE.4853..341S}
where we focus on this relation.

Section~\ref{sec:obs} describes the target, the observational setup, and the data reduction; Section~\ref{sec:results} contains our analysis and results; we conclude in Section~\ref{sec:conclusions}.

\section{Observations and data reduction} \label{sec:obs}

\begin{figure*}
\centering
\includegraphics[width=16.5cm]{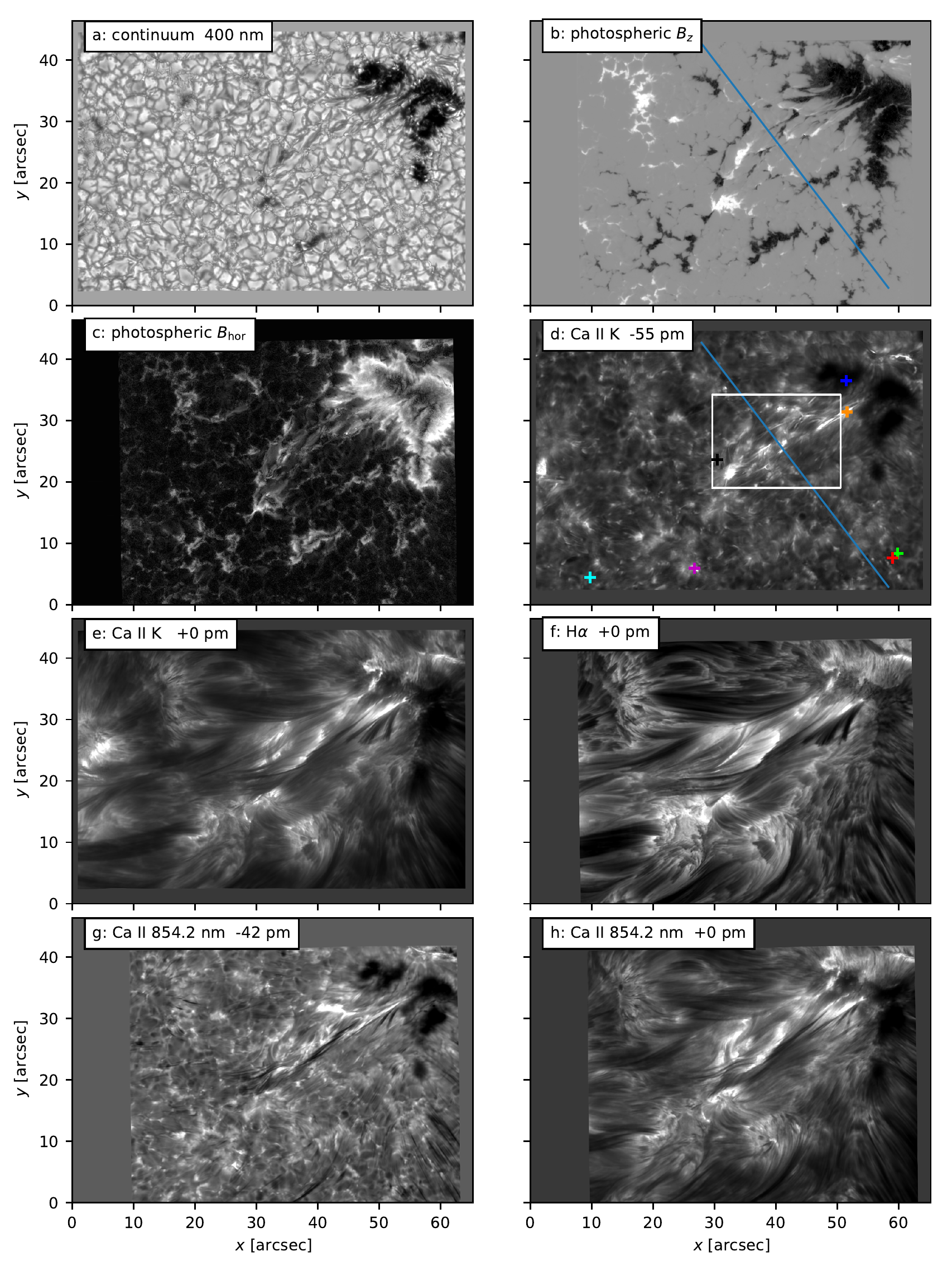}
\caption{Overview of observations taken at 09:44:37~UT. \textbf{a)} continuum at 400~nm;  \textbf{b)} $B_z$ as inferred from a Milne-Eddington inversion of the \FeI\ 603.1~nm and 630.2~nm lines\edt{, the color scale ranges from -2.0 kG to 1.4~kG}; \textbf{c)}  horizontal magnetic field from the same inversion\edt{, the color scale ranges from 0 kG to 1.3~kG}; \textbf{d)}  \CaIIK\ image in the \konev\ minimum of the average line profile; \textbf{e)}   \CaIIK\ at nominal line center;  \textbf{f)} \Halpha\ at nominal line center;  \textbf{g)}  blue wing of the \CaII\ 854.2 nm line; \textbf{h)}  \CaII\ 854.2 nm at nominal line center. The color scales are clipped to exclude the 0.5\% darkest and brightest pixels.  The colored plusses in panel d) indicate the locations for which line profiles are shown in Fig.~\ref{fig:overview_profiles} and Fig.~\ref{fig:classification_example}; the location of the $xy$ subfield shown in Fig.~\ref{fig:xy_inv} is indicated by a white box; the location of the timeslice shown in Fig.~\ref{fig:st_inv} with a blue line. An animated version of this figure showing the entire time sequence is available online.}
\label{fig:overview}
\end{figure*}

\subsection{Target and observational setup}

Our target was the active region NOAA 12593, which started emerging on 2016-09-19 around 00:00~UT. We observed it  with the SST 
on 2016-09-19 between 09:31:29~UT and 09:57:03~UT, using the CRISP
\citepads{2008ApJ...689L..69S}
and CHROMIS
\citepads{chromis2017}
instruments. The active region was located close to disk center, with heliocentric coordinates of the center of the field-of-view (FOV) at the beginning of the observations $(x,y)=(-68,-6)$~arcsec, and an observing angle  $\mu=1.00$.

The CRISP instrument observed the \FeI\ 630.1~nm line with 9 equidistant wavelength positions between $\pm19$~pm and the 603.2~nm line with 7 wavelength positions between $-28$~pm and $+8$~pm with full Stokes polarimetry. The \Halpha\ line was sampled in 16 wavelength positions non-equidistantly spaced between $-210$~pm and $+140$~pm from the line core. CRISP also observed the \CaII\ 854.2~nm line with 21 positions with two wavelength positions at $\pm170$~pm from the line core, and 19 positions spaced evenly between $-76.5$~pm and $+76.5$~pm with full Stokes polarimetry. The total cadence of this sequence was 36.6~s.

CHROMIS was used to observe the core of the \CaIIK\ line as well as a single wavelength point in the continuum at 400.0 nm. The \CaIIK\ line was sampled using 39 points: two wavelength positions at $\pm133$~pm from the line core, and 37 positions spaced evenly between $-70.5$~pm and $+70.5$~pm, with 3.8~pm between the positions. The cadence was 14.5~s, except for a few scans around the 100th repeat of the sequence where it rose to 16~s owing to a computer problem. 
The spectral PSF of  CHROMIS at 393~nm is 12~pm.

The data were post processed using Multi-Object Multi-Frame Blind Deconvolution
\citepads[MOMFBD,][]{2005SoPh..228..191V}
using the CRISPRED data processing pipeline
\citepads{2015A&A...573A..40D}, 
with modifications made as necessary to accomodate the CHROMIS data. The CRISP data were then carefully aligned with the CHROMIS data and resampled to the CHROMIS pixel scale of $0\farcs0375$. We estimate that the alignment is better than $0\farcs05$. 

Because the CRISP data were obtained with a lower  cadence we interpolated the CRISP data to the CHROMIS cadence using nearest-neighbour interpolation in the time dimension. As a final step we performed an absolute intensity calibration of the data by scaling the data numbers in a quiet patch of the observations to the intensity of the atlas profiles from
\citetads{1984SoPh...90..205N}.

\subsection{Overview of observations}

\begin{figure}
\centering
\includegraphics[width=8.8cm]{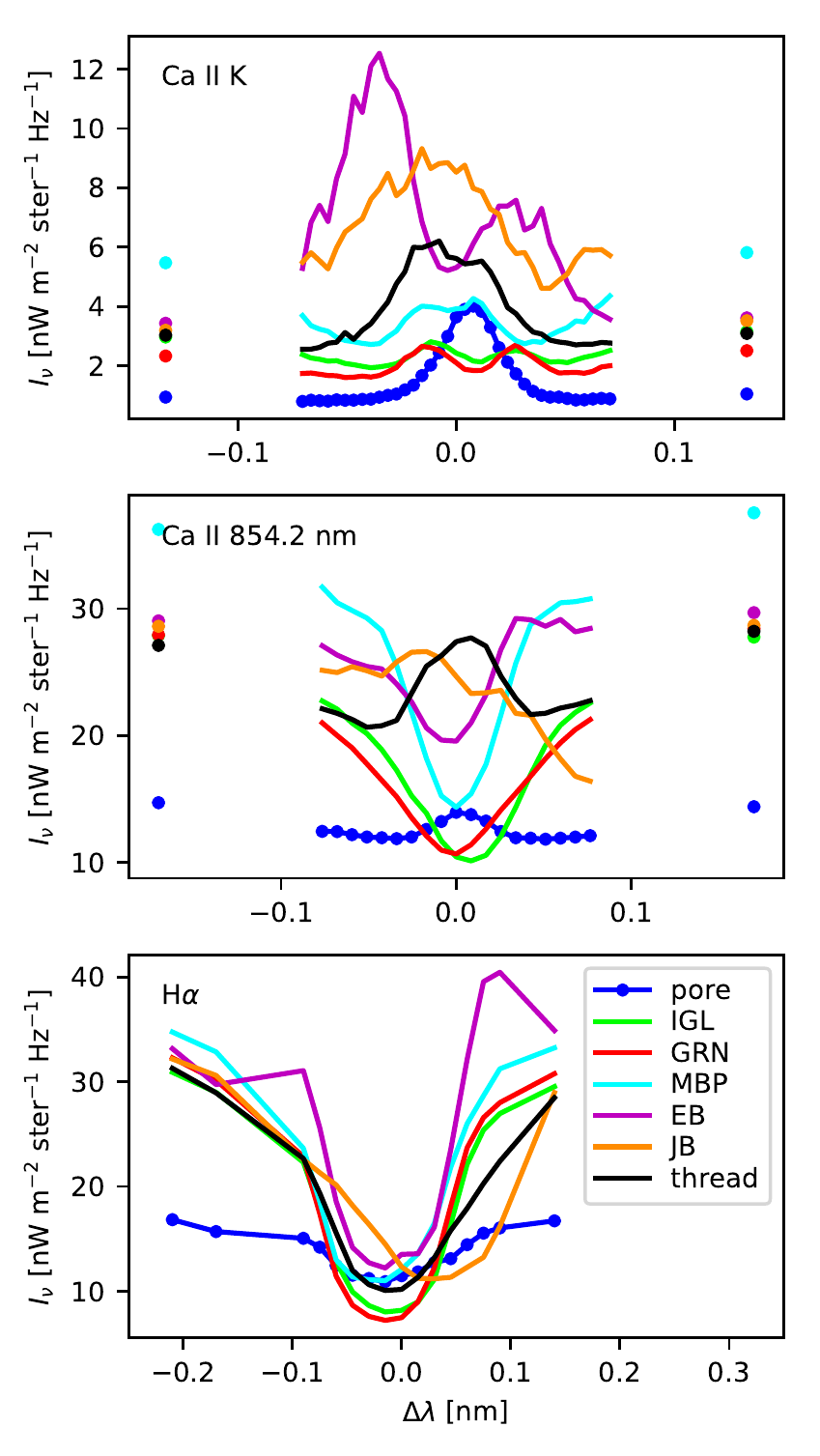}
\caption{Observed line profiles for the locations indicated with plus signs in panel d) of Fig.~\ref{fig:overview}. The color of the line profile matches the color of the plus sign. IGL: intergranular lane, GRN: granule center, MBP: magnetic bright point, EB: Ellerman Bomb, JB: jet base. The outlying wavelength points for the \CaII\ profiles are not connected by a line to avoid a false impression of the line profile shape. The filled circles in the pore profile shows the wavelength sampling of the observations.
}
\label{fig:overview_profiles}
\end{figure}

Figure~\ref{fig:overview} shows an overview of the observations. The CRISP and CHROMIS data do not have the same FOV, and the centers of their FOV are not exactly at the same location. Panel~a shows the continuum, with a group of large pores in the upper right and smaller pores scattered throughout the FOV. Panels~b and~c show the photospheric magnetic field (see Sec.~\ref{data_products}). The FOV shows a complex pattern of positive and negative magnetic flux density, with elongated granulation and extended patches of medium strength horizontal field indicating ongoing flux emergence along the line between $(x,y)=(30,14)$~arcsec and $(x,y)=(48,33)$~arcsec. The images in the wings of the \CaII\ lines (panels~d and~g) show enhanced brightness above the flux emergence. The line-core images (panels~e, f, and~h) show a complex structure of chromospheric fibrils, indicating the complex topology of the chromospheric magnetic field. We note that panels~e and~f show fan-shaped jets around $(x,y)=(50,30)$~arcsec. The accompanying movie of the entire time sequence display strong dynamics including fine bright threads seen in the blue wing of \CaIIK\ (panel~d).

Figure~\ref{fig:overview_profiles} shows example line profiles at the locations of the colored crosses in panel~d of Fig.~\ref{fig:overview}, and illustrates the wavelength sampling.
The line profiles show a large variation depending on the sampled structure.  Of particular interest are the \CaIIK\ profiles in the Ellerman bomb, the location from where the jets are launched (jet base), and the bright thread. The profiles have a high intensity in their line cores, and the cores are very wide compared to the more usual profiles in a granule, intergranular lane and the magnetic bright point. We also note that that the magnetic bright point has the highest intensity in the outermost points at $\pm0.133$~nm.

\subsection{Further data products} \label{data_products}

\begin{figure}
\centering
\includegraphics[width=8.8cm]{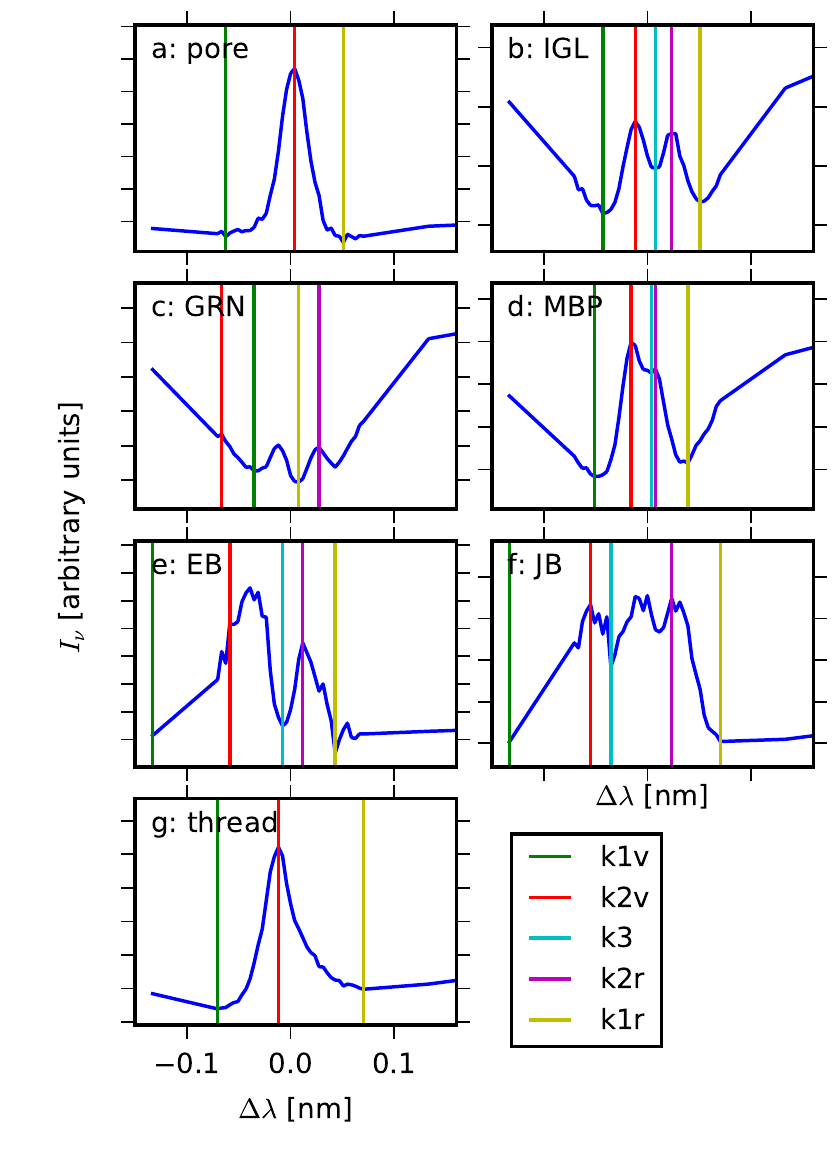}
\caption{Examples of the automated feature detection for the \CaIIK\ line profiles for different atmospheric structures. The vertical colored lines indicate the wavelength positions of features indicated in the legend. IGL: intergranular lane, GRN: granule center, MBP: magnetic bright point, EB: Ellerman Bomb, JB: jet base.}
\label{fig:classification_example}
\end{figure}

The entire sequence of \FeI\ data was inverted with a modified version of the Milne-Eddington inversion code presented in
\citetads{2015A&A...577A.140A}
operating in 1.5D mode, yielding estimates of the magnetic field vector in the photosphere. Example images of the observations and the inferred photospheric magnetic field are shown in Fig.~\ref{fig:overview}. 

We used an automated feature detection code to characterize the \CaIIK\ line profiles. This code tries to find the location of the \konev, \ktwov, \kthree, \ktwor, and \koner\ spectral features for a given line profile. Once it has found these locations it stores the intensity and wavelength of the features, and computes the full-width-at-half-maximum (FWHM) and full-width-at-quarter-maximum (FWQM) by finding the wavelengths  of the two outmost intersections of the line profile between \konev\ and \koner\ with the line 
\be
I = \min(I_\mathrm{K1v},I_\mathrm{K1r}) + a \left( \max(I_\mathrm{K2v},I_\mathrm{K2r})-\min(I_\mathrm{K1v},I_\mathrm{K1r}) \right),
\ee
where $a=0.5$ for the FWHM and $a=0.25$ for the FWQM. The difference of the two wavelengths is the width measure. The code also stores the average of these wavelengths, which measures the Doppler-shift of the emission peak at the quarter and half maximum. Not all profiles show a central emission core or a \kthree\ minimum. Such profiles are flagged as {\em pure absorption} (no \kone\ and \ktwo) or {\em single-peaked} (no \ktwor\ and \kthree). In Fig.~\ref{fig:classification_example} we show the results of the feature finder for the pixels indicated with plus-signs in Fig.~\ref{fig:overview}. The code correctly fits the features in panels a, b, d, and g, but misidentifies one or more features in panels c and e. The profile in panel f show three emission peaks with some additional wiggles, and the classification using \ktwov, \kthree, \ktwor\ is too simple. We note that because of the limited sampling of the line profile (see Fig.~\ref{fig:overview_profiles}) we  cannot determine the location of all features accurately, such as the locations of \konev\ in panels e and f. 
 Manual inspection of images of the various fitted properties show however that the misidentification rate is rather small, and typically happens for very bright and wide profiles with  multiple emission peaks such as in panel e and f of Fig.~\ref{fig:classification_example}.

\section{Results} \label{sec:results}

\begin{figure}
\centering
\includegraphics[width=8.8cm]{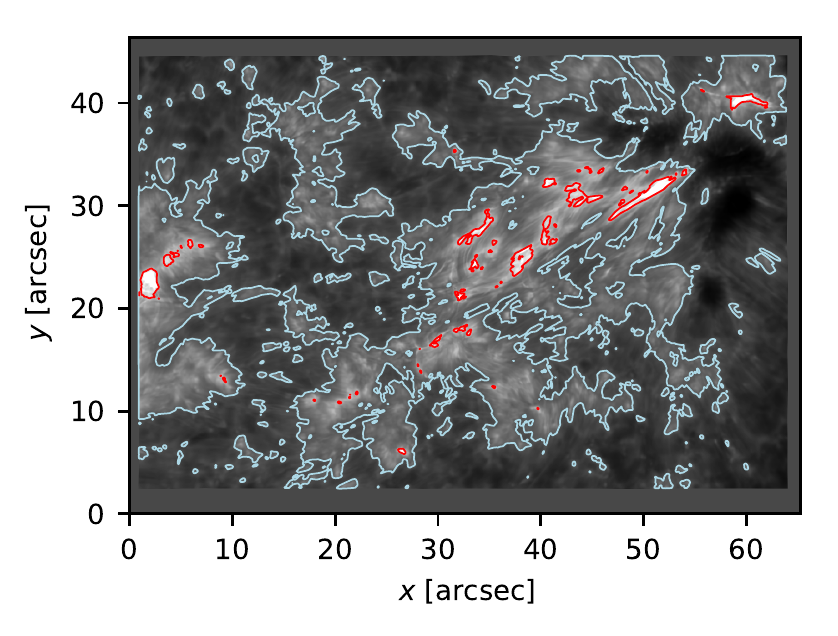}
\caption{Image of the wavelength-summed \CaIIK\ intensity taken at 09:44~UT. The light blue and red contours divide the image into low, intermediate and high brightness areas. An animated version of this figure showing the entire time sequence is available online.}
\label{fig:brightness_contours}
\end{figure}

\subsection{\CaIIK\ brightness and chromospheric heating} \label{subsec:brightness_and_heating}

The precise mechanisms that heat the solar chromosphere under various circumstances are not yet known and understood. These processes must act because the average chromosphere emits more radiation than predicted by radiative equilibrium models
\citepads[e.g.][]{1977ARA&A..15..363W},
even in the quiet Sun. A fundamental problem is that neither the local heating rate nor even the local radiative cooling rate can be measured directly. The local net radiative cooling rate is 
\be
\oint_\Omega \int_0^\infty \alpha_\nu \left( S_\nu - I_\nu \right) \, \dd \nu \,\dd \Omega,
\ee
with $\alpha_\nu$ the extinction coefficient, $S_\nu$ the source function, and $I_\nu$ the intensity, all as function of frequency $\nu$ and direction $\Omega$. What we measure is the emergent intensity:
\be
I_\nu(z=\infty) = \int_{z_0}^\infty \alpha_\nu  \, S_\nu \, \mathrm{e}^{-\tau_\nu} \, \dd z,
\ee
with $z_0$ a location sufficiently deep in the atmosphere so that the integrand is zero and $\tau_\nu$ the optical depth. The only way to get a quantitative estimate of the cooling rate from observations is to construct an atmosphere model that is consistent with the observed intensity $I_\nu$ and compute the radiative cooling in the  model. This is beyond the scope fo the current paper.

Here we take a simpler approach: We assume that the \CaIIK\ intensity, frequency-integrated over the inner wings and line core, correlates with the total radiative cooling along the line-of-sight, and thus with conversion of non-radiative and non-thermal energy into heat in order to sustain the radiative losses. This assumption is justified based on radiative transfer modelling that shows that increased \CaIIK\ emission indeed corresponds to elevated temperatures in the chromosphere in both 1D semi-empirical atmosphere models as well as those computed from radiation-MHD models
\citep{1981ApJS...45..635V,2012A&A...539A..39C}.

In order to make the connection between our spectrally-resolved data and total core emission we summed the intensities at all wavelength positions for each timestep in our \CaIIK\ data. We then selected two brightness thresholds at $1.05$  and 2.09 times the spatially and temporally averaged value of the data. The lower threshold divides the images into rather persistent dark and intermediately-bright areas. Inside the intermediate-brightness bin we see many instances of high brightness, some persisting over the whole time series, and some more intermittent, which fall above the 2.09 level. The selection of the thresholds is somewhat arbitrary, but serves adequately to \edt{roughly} separate the images into quiet-Sun-like areas, the flux-emergence area and smaller high-brightness events. We show this in Fig.~\ref{fig:brightness_contours} and the accompanying online movie.

In total, 66.8\% of the pixels fall into the low-brightness bin, 32.6\% in the intermediate bin and 0.60\% in the high brightness bin. The high-brightness pixels produce 3.3\% of the total integrated observed emission in the intermediate and high brightness bin (i.e., the area associated with the flux emergence). 
\edt{ Inspection of the high-brightness bin events shows that they are typically associated to cancellation of opposite polarity field in the photosphere and/or reconnection between horizontal and vertical field in the low chromosphere. They are spectacular, but appear to play only a minor role in the overall heating because they cover only a small fraction of the area.}
The exact percentages obviously depend on the threshold values, \edt{as well as the selection of the FOV. In our case some of the areas around the edges of the FOV with intermediate brightness do not contain vigorous flux emergence.}
\edt{However, even} with different threshold values and FOV selection the question of chromospheric heating is then: what caused the overwhelming majority of the emission in the emerging flux region?

\subsection{\CaIIK\ profile properties and total brightness }

\begin{figure*}
\centering
\includegraphics[width=17cm]{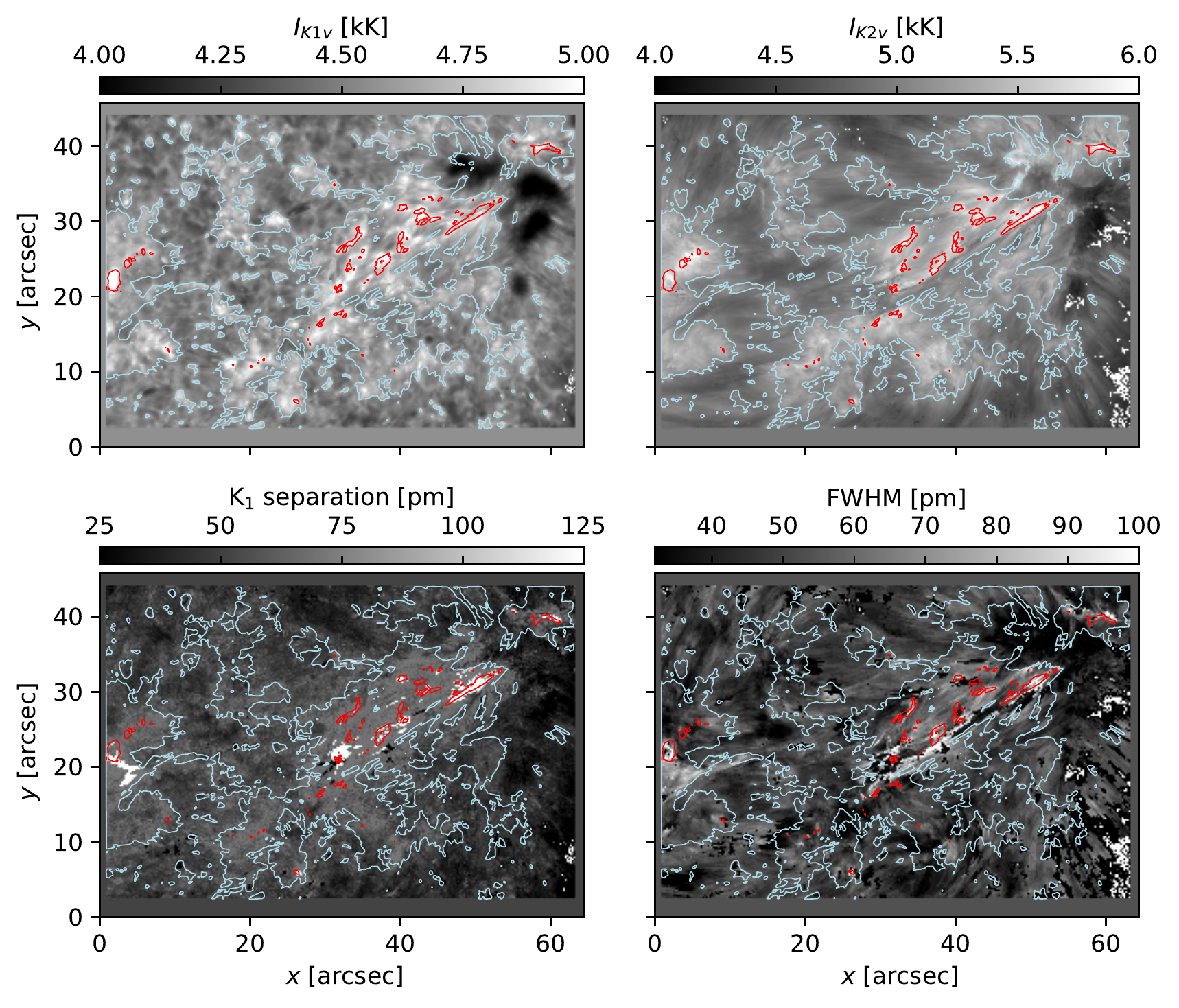}
\caption{Properties of the \CaIIK\ line profiles at 09:44:37~UT. Top row: Intensities expressed as radiation temperatures in \konev\ and \ktwov. Bottom row: wavelength separation of the \kone\ minima and full width at half maximum of the central emission peak. The brightness range of all images has been clipped to enhance contrast. The light blue and red contours divide the image into low, intermediate and high brightness areas of the corresponding wavelength-summed \CaIIK\ image.}
\label{fig:overview_peaks}
\end{figure*}

\begin{figure*}
\centering
\includegraphics[width=17cm]{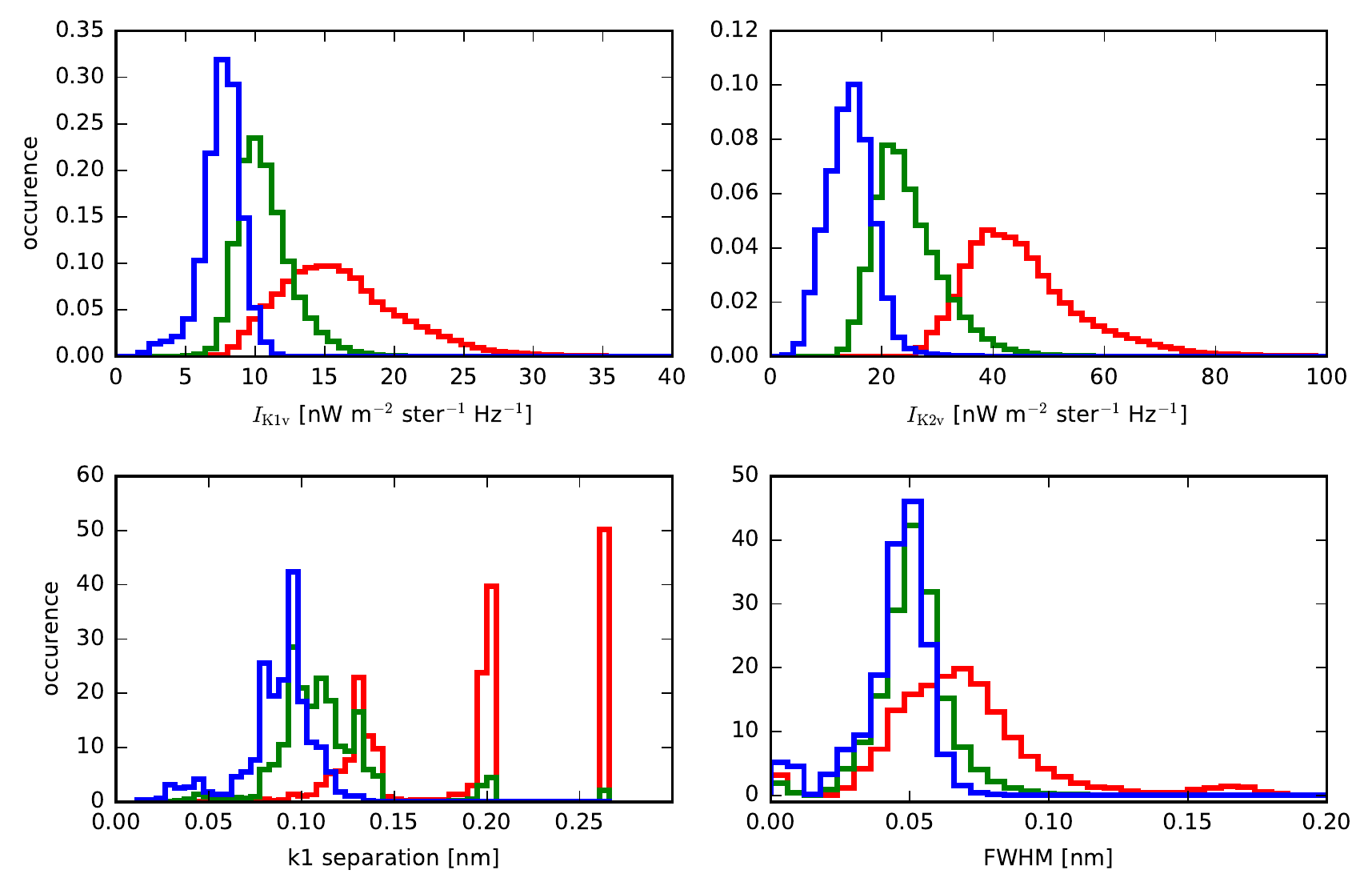}
\caption{Histogram of \CaIIK\ line profile features per brightness bin. Blue: low brightness, green: intermediate brightness, red: high brightness. Upper-left: intensity of the \konev\ feature. Upper-right: inensity of the \ktwov\ features. Lower-left: $\lambda_{\mathrm{K1r}} - \lambda_{\mathrm{K1v}}$, the wavelength separation between the K$_1$ minima. The three-peaked structure of the green and red distributions are caused by the sparse wavelength sampling furthest away from the line core. Lower-right: the full-width at half maximum of the emission peak between the K$_1$ minima. \edt{The histograms are normalized so that the sum of all bins times the bin width adds up to one.}}
\label{fig:plot-level-hist}
\end{figure*}

 In Fig.~\ref{fig:overview_peaks} we show images of  the intensity of \konev\ and \ktwov\ as well as the wavelength separation of the \kone\ minima and the FWHM of the central emission peaks. The pores show the lowest \konev\ intensity, with $\Trad = 3.89$~kK, and the highest intensities are in the bright patches associated with the flux emergence where $\Trad$ can be as high as 5.78~kK. The  \ktwov\ intensities are higher, between 4.02~kK and 7.04~kK, with again the lowest values in the pores, and the highest values in the region of the ongoing flux emergence.
The \kone\ separation and FWHM roughly fall between 30~pm and 170~pm, with lower values for the FWHM. The lowest width values are associated with pores and areas with weak photospheric magnetic field (compare Fig.~\ref{fig:overview}). The highest widths are located again where the flux emergence is taking place. The black pixels around this area are locations where the classification algorithm failed in correctly determining one or more of the spectral features.

In Figure~\ref{fig:plot-level-hist} we show histograms of the \CaIIK\ profile properties separated for each brightness bin. These histograms were created from the entire time-series. The mean \konev\ and \ktwov\ brightness increases when going from the low-brightness to the high-brightness bin. The same holds for the \kone\ separation. We note that the peaks in the histogram at 0.20~nm and 0.26~nm are a consequence of the large wavelength difference between the densely sampled line core and the two outlying outer wavelength points. The peaks are formed of those pixels where one or both of the minima are located in the outer points.
The FWHM of the central emission peak behaves differently: the distributions for the low and intermediate brightness bins are very similar, and only the high-brightness bin shows a clear increase in FWHM.

\subsection{\CaIIK\ intensity and chromospheric magnetic field}

\begin{figure*}
\centering
\includegraphics[width=17cm]{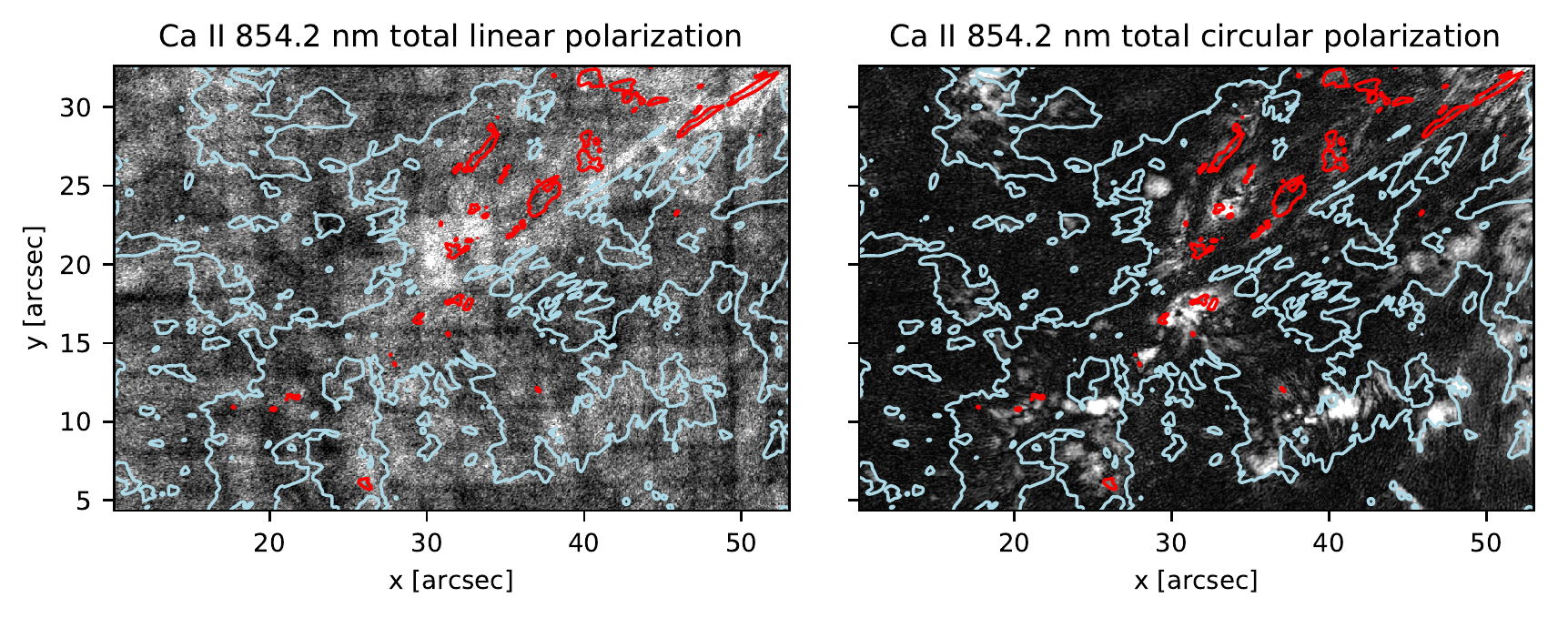}
\includegraphics[width=7cm]{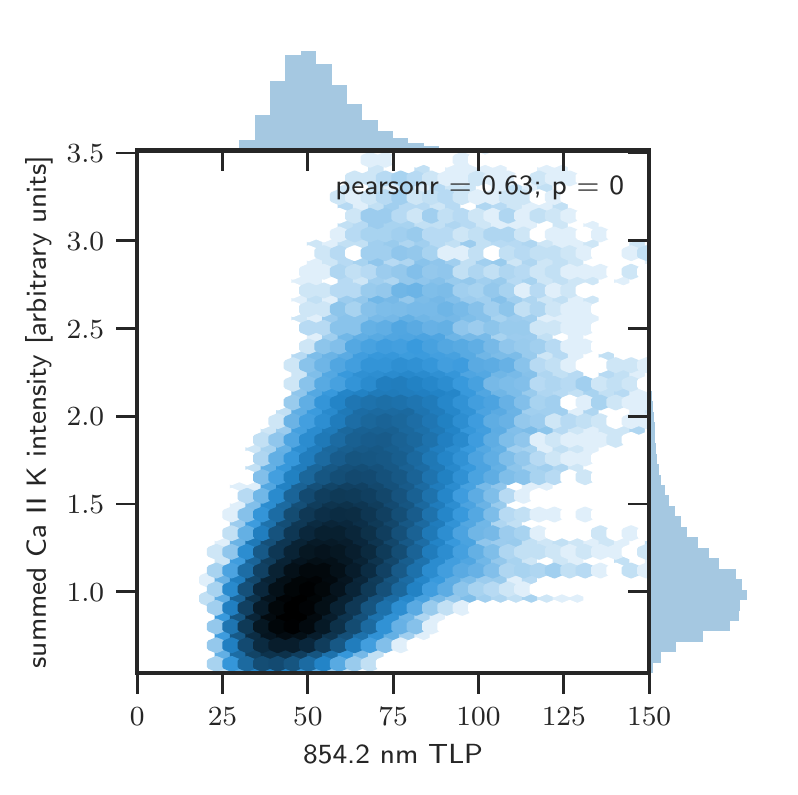}
\includegraphics[width=7cm]{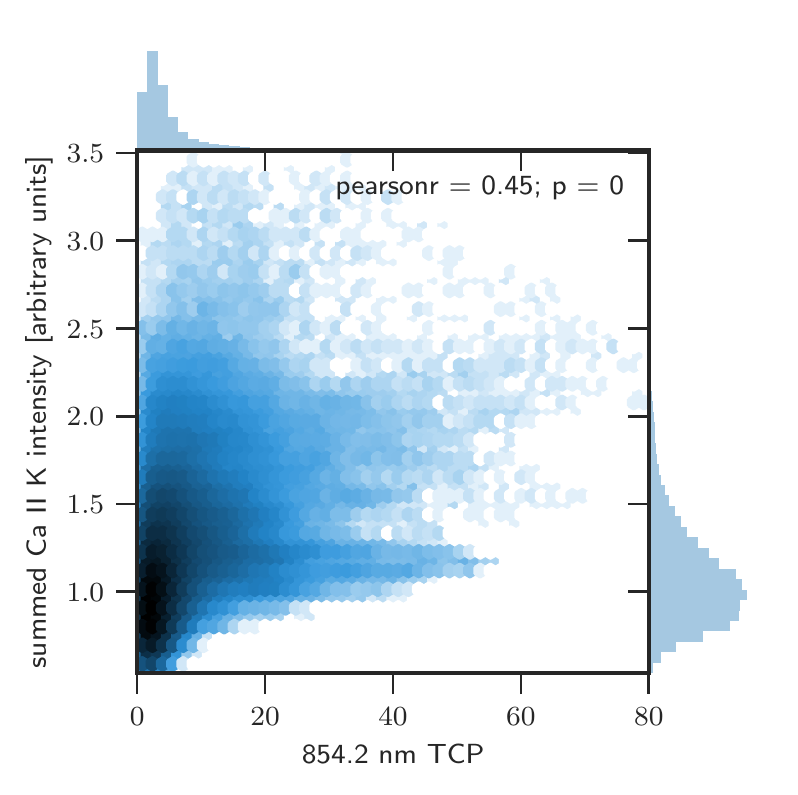}
\caption{\CaIIK\ summed intensity and polarization signal in \CaII~854.2~nm \edt{at 09:44~UT}. Upper panels: Maps of the total linear and circular polarization in the \CaII~854.2~nm line with the wavelength-summed \CaIIK\ brightness contours defined in Sec.~\ref{subsec:brightness_and_heating} overplotted. The brightness range of the images has been clipped to enhance contrast. Lower panels: joint probability distributions of the total linear and circular polarization in the \CaII~854.2~nm line and the wavelength-summed \CaIIK\ intensity \edt{computed from the data shown in the upper panels}. The Pearson correlation coefficient $r$ and the $p$ value are indicated at the top of both panels. The color of the hexagons is proportional to the logarithm of the number of points in each bin, with darker color indicating more points.}
\label{fig:brightness_corrs}
\end{figure*}

We now turn our attention to the relation between the summed \CaIIK\ intensity and the chromospheric magnetic field. The datasets contain full-Stokes observations in the \CaIIIR\ line. The polarimetric accuracy of this data is of the order of  $\sim3\times10^{-3}$ and meaningful signal in Stokes $Q$ and $U$ is only obtained in a small fraction of the pixels, mainly located close to the pores. The Stokes $V$ signals are stronger but also here we do not detect signal throughout the field-of-view. In order to boost the signal-to-noise ratio we therefore opted to compute the total linear polarization:
\be
\mathrm{TLP} = \sum_{i=1}^{21} \sqrt{Q_i^2 + U_i^2},
\ee
where the sum is taken over all 21 observed wavelength positions in the \CaIIIR\ line. \edt{Including  the far wing wavelengths does not add photospheric contamination because the polarization signal there is very weak.} We also computed the integrated circular polarization:
\be
\mathrm{TCP} = \sum_{i=6}^{16} |V_i| 
\ee
where the sum is over the observed wavelength positions less than 42.5~pm from nominal line center. The wavelength range is constrained to the line core because Stokes $V$ shows noticeable photospheric signal in the line wings.

These quantities do not provide quantitative information about the horizontal and vertical components of the magnetic field, but do provide qualitative information: Stronger TLP (TCP) generally means stronger horizontal (vertical) magnetic field strengths around the height where the response function peaks. This height is located roughly between $-4 < \log{\tau_{500}} < -5$ in the 1D semi-empirical FALC atmosphere model
\citepads{2016MNRAS.459.3363Q}.
In this flux emergence region this optical depth might be lower owing to the elevated density compared to the quiet Sun
\citepads{2017MNRAS.472..727Q}.

The upper row of Fig.~\ref{fig:brightness_corrs}  shows maps of TLP and TCP with wavelength-summed \CaIIK\ brightness contours overplotted at 09:43:54~UT. This time was chosen to minimize the time difference between the \CaIIK\ and \CaIIIR\ scans owing to the different cadence of the CRISP and CHROMIS observations. 

The \CaIIIR\ line is not only sensitive to the Zeeman effect, but also to atomic level polarization due to anisotropy in the radiation field and depolarization by the Hanle effect
\citepads{2010ApJ...722.1416M,2015ApJ...801...16C,2016ApJ...826L..10S}.
Three-dimensional radiative transfer calculations computed with a 3D radiation-MHD simulation as input atmosphere indicate that atomic level polarization and the Hanle effect generate linear polarization signals at the $10^{-3}$ level. These signals dominate in quiet Sun areas of the model atmosphere, while in areas with strong horizontal fields the linear polarization is dominated by the Zeeman effect
\citepads{2016ApJ...826L..10S}.
This gives us reason to believe that our conclusions are not affected by scattering polarization. 

The TLP map structure is also not caused by cross-talk from Stokes $I$. \edt{While the TLP map
shows a rough correlation with the wavelength-summed CaII 854.2 nm intensity,} sharp individual bright and dark structures structures in the intensity image are not present in the TLP map, which would be the case if the cause was cross-talk.

The TLP panel shows a pattern of dark square grid lines. These are the edges of the MOMFBD image restoration patches. At the edges the final signal is composed of a linear combination of both patches, lowering the noise, and thus appearing dark in the image. The TLP appears concentrated within the blue contours, with the highest values in the upper-right corners close to the pores (cf. Fig.~\ref{fig:brightness_contours}. While some of the highest TLP values fall within the red high-\CaIIK-brightness contour, there are many high values that do not. 


The TCP panel shows more intermittent structure, with the strongest signal above the photospheric magnetic elements. Almost all signal is contained within the blue contours.

The lower panels show joint probability distributions of the total linear and circular polarization in the \CaIIIR\ line and the wavelength-summed \CaIIK\ intensity. The TLP panel shows that high TLP correlates with high \CaIIK\ brightness, even though the distribution has a fairly large spread. The TCP panel shows a more ambiguous result. 
The correlation coefficient is still relatively high at $r=0.45$, but the shape of the distribution shows considerably more structure than the TLP panel. The most notable case is the horizontal arm at a \CaIIK\ intensity of $1.2$.

We interpret these results as follows: The flux emergence produces a multitude of low-lying magnetic field lines in the lower chromosphere. These field lines are rooted in photospheric magnetic elements. Above these elements the magnetic field still has an appreciable vertical component, which is depicted in the TCP image, but further away from the magnetic elements the field is mainly horizontal. The horizontal field in the low chromosphere where the TLP signal is formed appears space-filling. The correlation between TLP and \CaIIK\ intensity implies a heating function that scales with magnetic field strength. Because neither diagnostics are linear (unknown dependence for \CaIIK\ brightness and chromospheric heating, while TLP scales as $B_\mathrm{hor}^2$) we cannot comment on the functional dependence of heating as function of horizontal field strength.

\subsection{Non-LTE Inversions}

\begin{figure*}
\centering
\includegraphics[width=17cm]{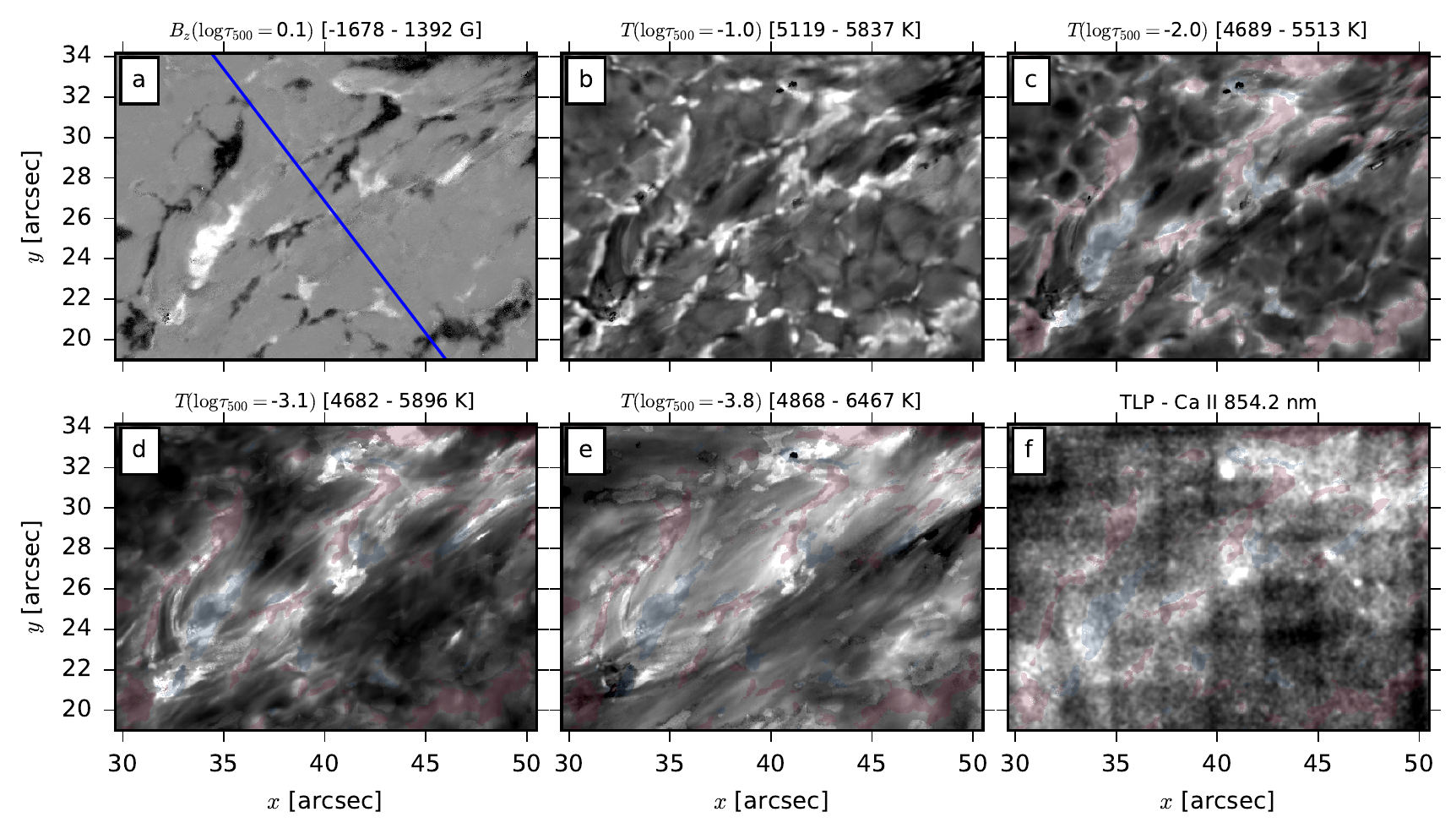}
\caption{Results from the non-LTE inversion of the \CaIIK, \FeI, and \CaIIIR\ lines in a 2D field-of-view at 09:43:54~UT. Panel a shows the vertical magnetic field in the photosphere, while panels b\,--\,e show the inferred temperature at different 500 nm continuum optical depths, as specified above each panel. Panel f displays the total linear polarization in the \CaIIIR\ line. Panels d\,--\,f show the photospheric vertical magnetic field in semi-transparent red and blue colors for comparison.
 The overlap with the $st$-slice shown in Fig~\ref{fig:st_inv} is indicated by the blue line. The brightness of each panel   is clipped at the brightest and darkest 0.5\% of the pixels. The range of the grey scales is indicated between square brackets. The field-of-view used for the inversion is indicated in panel~d of Fig.~\ref{fig:overview}. }
\label{fig:xy_inv}
\end{figure*}

\begin{figure*}
\centering
\includegraphics[width=17cm]{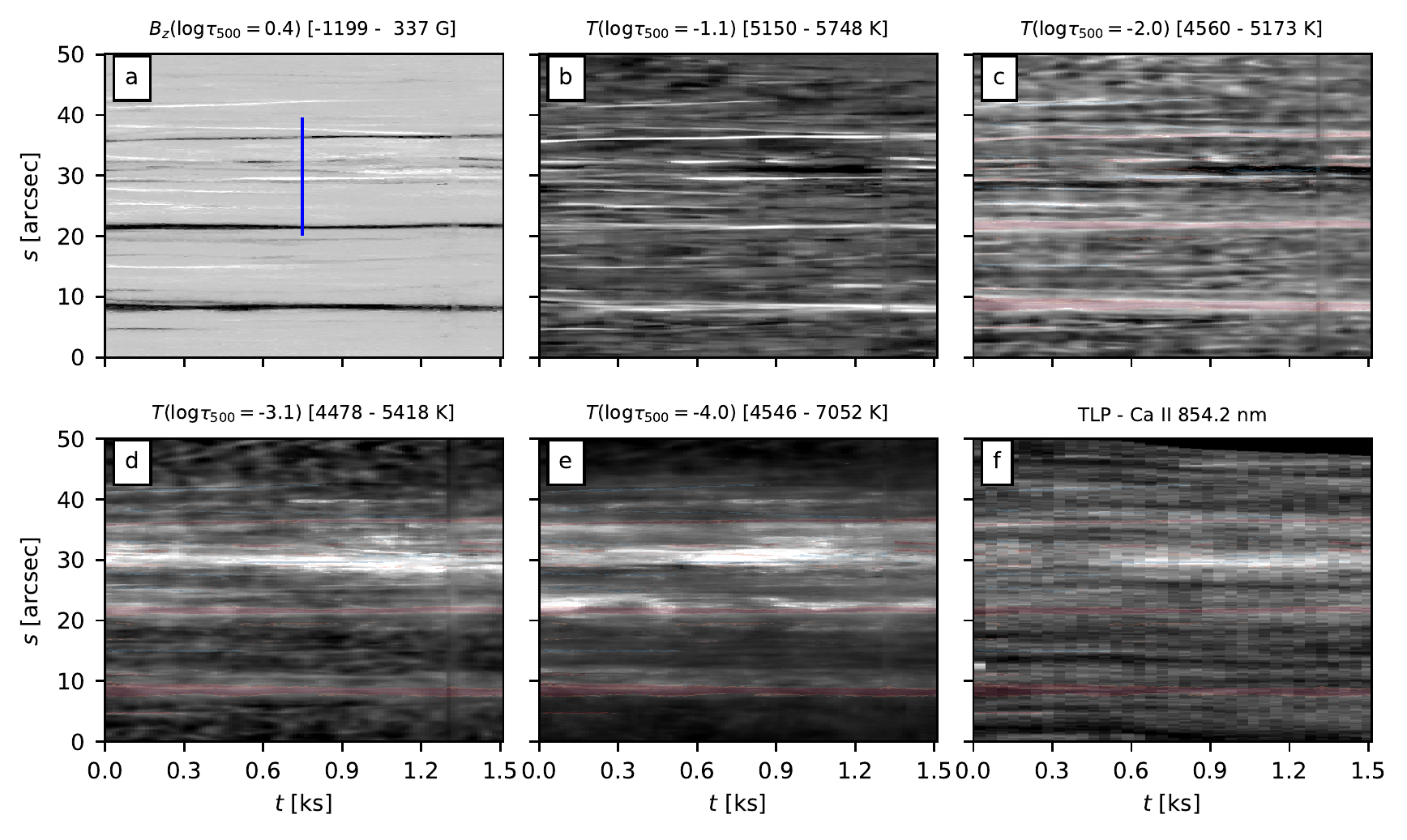}
\caption{Results from the non-LTE inversion of the \CaIIK, \FeI, and \CaIIIR\ lines in a time-slice indicated by the diagonal line in panel~d of Fig.~\ref{fig:overview}. The upper left panel shows the \edt{vertical} magnetic field in the photosphere, while the panel \edt{c--f} show the inferred temperature at different 500 nm continuum optical depths, as specified above each panel. \edt{Panel f shows the total linear polarization signal in the \CaIIIR\ line.} The brightness of each panel  is clipped at the brightest and darkest 0.5\% of the pixels. The range of the color scales is indicated between square brackets. Panels \edt{c--f} show the photospheric magnetic field in semi-transparent red and blue for comparison. The overlap with the $xy$-region shown in Fig.~\ref{fig:xy_inv} is indicated by the blue line in the upper left panel. } 
\label{fig:st_inv}
\end{figure*}

\begin{table}
\caption{Number of nodes used in the inversions.}\label{table:inv_params} 
\centering
\begin{tabular}{lccc}
\hline\hline
 Quantity & $xy$-image & $st$-slice \\ 
\hline
$\min( \log \tau_{500})$ & -8.2 & -7.8\\
$\max(\log \tau_{500})$ & 0.6 & 0.4 \\
$T$ & 10& 9 \\
$v_z$ & 9 & 5 \\
$v_\mathrm{turb}$ & none & 4\\
$|\vec{B}|$ &2 & 3\\
inclination & 2& 3\\
azimuth & 1 & 1\\
\hline
\end{tabular}
\end{table}

As a final step in the analysis of the data we estimate the 3D structure of the temperature in the emerging flux region through non-Local Thermodynamic Equilibrium (non-LTE) inversion of our observations.

We perform  the inversion using the STiC inversion code developed by
\citetads{2016ApJ...830L..30D}.
This code uses the RH code
\citepads{2001ApJ...557..389U}
as the radiative transfer engine. It assumes that each pixel is an independent plane-parallel atmosphere, includes the effect of partial redistribution (PRD) in moving atmospheres following
\citetads{2012A&A...543A.109L},
and can solve the non-LTE radiative transfer problem for multiple atomic species and lines simultaneously. 

The inversion code produces as output for each pixel a 1D atmosphere (temperature, vertical velocity, magnetic field vector and optionally microturbulence) on a $\log \tau_{500}$ height scale. The inversion code assumes hydrostatic equilibrium and a smooth variation of each parameter between the nodes.

Here we used the \CaIIK\ line, the \CaII\ 854.2~nm line, the \FeI\ 630.1~nm and 630.2~nm lines and the continuum at 400.0~nm simultaneously. We treated the \CaII\ atom in non-LTE with the H\&K lines in PRD. The \FeI\ lines are treated assuming LTE. We inverted the entire time series for all pixels along the blue line indicated in Fig.~\ref{fig:overview}, as well as a 2D field-of-view (white rectangle in Fig.~\ref{fig:overview})  in a single time step at 09:43:54~UT. 

Owing to the different cadence of CHROMIS (14.5~s) and CRISP (36.6~s), the maximum time difference between different spectral line scans that are inverted together can be up to 18~s. For the time-dependent inversion this is an unavoidable limitation. The time step for which we invert a 2D patch was chosen so that the time difference between the observation of the central wavelength of \CaIIK\ and \CaII~854.2~nm was 0.9~s, to minimise the inconsistencies of the multi-line inversion caused by the possibly fast temporal evolution in the chromosphere.

The $xy$-image and the $st$-slice inversion were performed with slightly different setups, which are summarized in Table.~\ref{table:inv_params}. 

The non-LTE inversion technique is a powerful method to derive atmospheric quantities. However, it has a number of limitations: The inversion assumes that each pixel is a plane-parallel atmosphere. Three-dimensional radiative transfer is however important in  the cores of the \CaII\ lines
\citepads{2009ASPC..415...87L,2009ApJ...694L.128L,bjoergen2017},
and its neglect might lead to errors in the upper chromosphere of the inferred  atmosphere models. 

The strongly scattering nature of these lines lead to a decreased contrast in the retrieved temperature maps. Test inversions with the \CaIIIR\ line indicate that small pockets of material cooler than $\sim$4000~K will typically not be retrieved 
\citepads{2012A&A...543A..34D}.

The inversions that we performed here assume a smooth variation of atmospheric parameters through spline interpolation in between the nodes
\citepads{1992ApJ...398..375R}.
While nodes can in principle be placed arbitrarily close together in $\log{\tau_{500}}$-space, in practice placing them equidistantly is the most practical approach when inverting large numbers of pixels. Our results can therefore never represent sharp vertical gradients in, for example, magnetic field strength and temperature that we expect to be present in the solar atmosphere. 

There is also a degeneracy between fitting line widths with microturbulence or temperature 
\citepads[cf. Figs. 4 and 5 of][for an example in the \ion{Mg}{II} h\&k lines, which have a similar formation mechanism as the \CaIIHK lines.]{2015ApJ...809L..30C}.
In test computations we find that the latter effect can lead to temperature differences up to 1000~K in the inferred atmospheres when performing inversions with and without microturbulence. 

Finally, the CRISP and CHROMIS data have different spatial resolution, and each wavelength position is observed at a different time, so that the spectrum fed into STiC is not strictly cotemporal, but can, in this case have a time difference up to 16 s between wavelength positions, which will affect results obtained for structures that evolve faster than that in time.

Despite these caveats, inversion of chromospheric lines is an excellent tool to build a spatially and temporally resolved picture of the solar atmosphere. It is a powerful method including much of the relevant physics, and is unbiased by human judgement. 

Figure~\ref{fig:xy_inv} shows the results for the inversion of the 2D field-of-view. The most vigourous flux emergence is occurring in a diagonal band from $(x,y) = (30,20)$ arcsec  to $(x,y) = (48,34)$ arcsec. 
 It shows the spatial structure of the temperature at different heights in the atmosphere, together with the photospheric vertical magnetic field for context. At $\log{\tau_{500}}=-1.0$ we see granulation and the heating in the magnetic elements. At $\log{\tau_{500}}=-2.0$ we see some reversed granulation, and the heating of the magnetic elements is more extended owing to the fanning out of the magnetic field. 
 
 At $\log{\tau_{500}}=-3.1$ the scene changes qualitatively. Strong increases in temperature start to appear that are not located on top of the photospheric field concentrations. In addition, a number of thin, thread-like regions of increased temperature are present that span between regions of opposite photospheric magnetic field polarity. At  $\log{\tau_{500}}=-3.8$ the temperature increase becomes space-filling in the emerging flux region, with typical temperatures around 6~kK. This behaviour is consistent with the usual picture of expansion of the magnetic field with height to form a magnetic canopy.
 
 Above $\log{\tau_{500}}=-4.0$ the inversion result is dominated by systematic errors and a lack of sensitivity, and therefore we do not show it here. Instead we show the cospatial \CaIIIR\ TLP map in panel f. Even though it is noisy, it shows a clear correlation with the $T(\log{\tau_{500}}=-3.8)$ in panel e.

In Figure~\ref{fig:st_inv} we display the time variation of the inferred atmosphere along a cut through the emerging flux region. The location of this cut is indicated by the blue line in Fig.~\ref{fig:overview}; the figure format is otherwise the same as for Fig.~\ref{fig:xy_inv}. We see a rather similar behaviour: at $\log{\tau_{500}}=-1.1$ the photospheric flux concentrations appear bright; higher in the atmosphere the heating becomes more spread out, and is not co-spatial with the photospheric flux elements. Around $\log{\tau_{500}}=-4$ the heating in the emerging flux region is rather homogenous, and the sites with the highest temperatures are mainly located  between  the photospheric flux elements. There is relatively little time variation in the temperature, indicating heating processes that are rather persistent in time.

\begin{figure*}
\centering
\includegraphics[width=7cm]{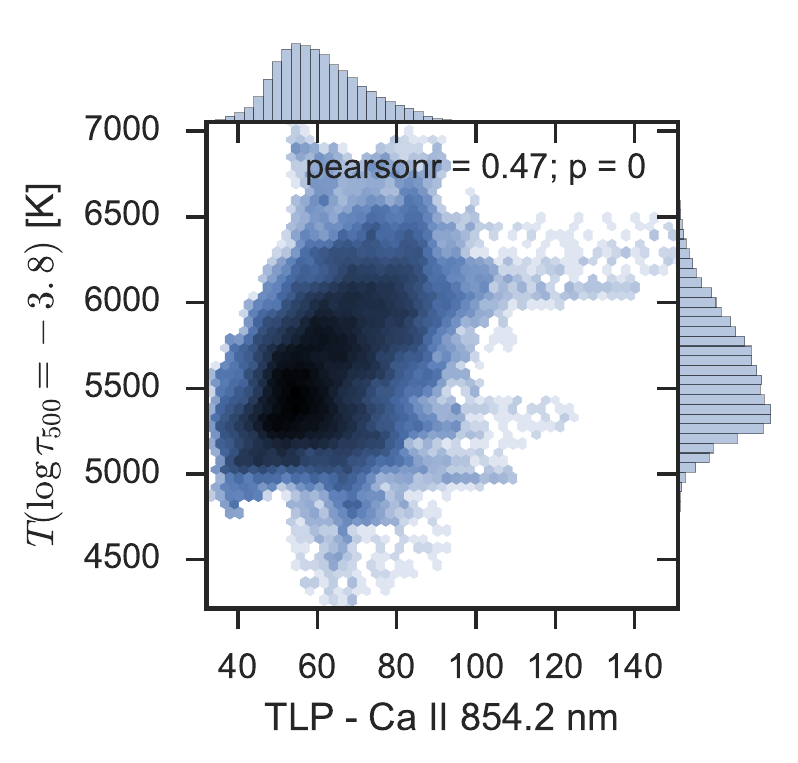}
\includegraphics[width=7cm]{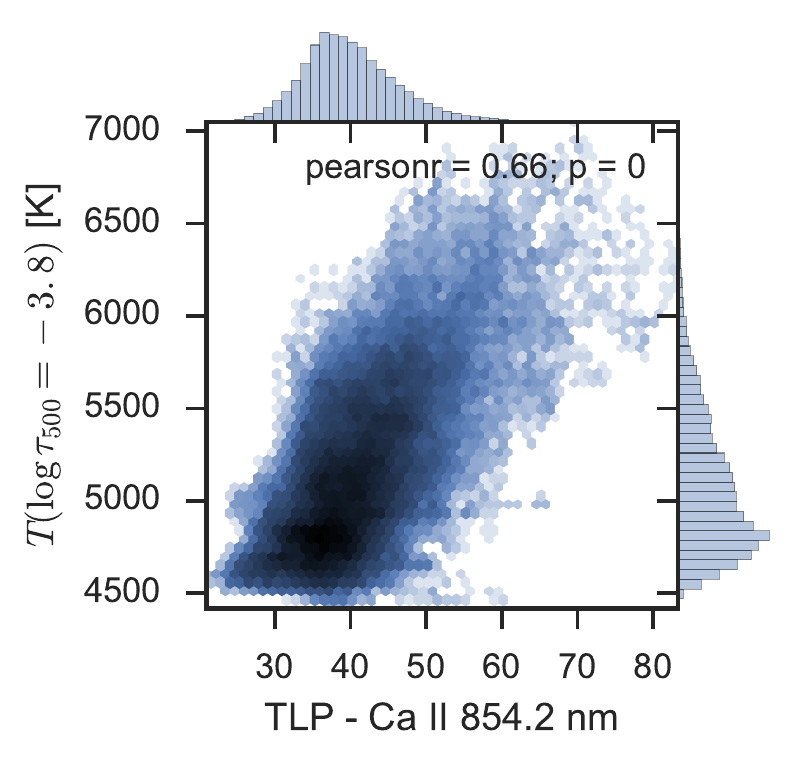}
\caption{Correlations between inferred gas temperature at $\log{\tau_{500}}=-3.8$ and the total linear polarization in the \CaIIIR\ line. Left-hand panel: for the $xy$-image shown in Fig.~\ref{fig:xy_inv}. Right-hand panel: for the $st$-slice shown in Fig.~\ref{fig:st_inv}. The Pearson correlation coefficient $r$ and the $p$ value are indicated at the top of both panels. The color of the hexagons is proportional to the logarithm of the number of points in each bin, with darker color indicating more points.}
\label{fig:TLP_T_corrs}
\end{figure*}

Finally, in Fig.~\ref{fig:TLP_T_corrs} we show the correlation between the inferred gas temperature at $\log{\tau_{500}}=-3.8$ and the total linear polarization in the \CaIIIR\ line. Both distributions have some outlying arms owing to noise and artefacts. but the bulk of the distributions follows a clear correlation, which is corroborated by the correlation coefficients.

\section{Summary and Conclusions} \label{sec:conclusions}

We present imaging spectropolarimetric observations of active region NOAA 12593. The chromosphere  above the flux emergence sites as seen in wavelength-summed \CaIIK\ intensity is persistently brighter than the surrounding atmosphere where no flux emergence is taking place.

We divide the excess brightening into a spatially diffuse, persistent-in-time, intermediate brightness part and a spatially localized and more intermittent high-brightness part. The intermediate brightness bin is the dominant contributor to the overall brightness of the flux emergence region. We show that the excess brightness in the intermediate bin is caused by increased \kone\ \edt{(forming around the temperature minimum)} and \ktwo\ \edt{(forming in the low chromosphere)} intensity, as well as an increase in the \kone\ separation compared to the low-brightness bin, but the FWHM of the emission peak is not increased. This indicates that the excess heating in the chromosphere occurs already deep in the chromosphere, around the temperature minimum and just above it. The lack of increase in FWHM might indicate that the upper chromosphere is not as strongly affected by the heating. Alternatively, the \CaIIK\  \edt{FWHM} might just not be sufficiently sensitive to temperature. It would be interesting to repeat this analysis with \ion{Mg}{II}\,h\&k data taken with the Interface Region Imaging Spectrograph
\citepads[IRIS, ][]{2014SoPh..289.2733D},
because those lines have higher opacity and are sensitive to temperatures at somewhat larger heights.

Analysis of the total linear polarization of the \CaIIIR\ line reveals that the emerging flux region is likely permeated with space-filling horizontal magnetic fields. The TLP correlates strongly with the wavelength-summed \CaIIK\ intensity, much more so than the TCP, which traces vertical field in the chromosphere. This result indicates that the chromospheric heating rate increases with increasing horizontal field strength.

Our multi-line, multi-atom, non-LTE inversions confirm that the excess brightening observed in \CaIIK\ indeed corresponds to an increased temperature at the temperature minimum and above. The inverted $xy$ image shows instances of thin threads of increased temperature around $\log \tau_{500} = -3.1$. We conjecture that these are the signature of heating by current sheets as seen in numerical simulations.
\citepads{2005ApJ...618L.153G,2015ApJ...811..106H}.
In our observations the threads appear around the height of the temperature minimum. This might be caused by a strong overlying canopy of previously emerged magnetic field that forces the interaction region where the currents form to lower heights than in the simulations. 

Around $\log \tau_{500} = -4.0$ the temperature in the flux emergence region appears more homogenous, both in space and time, even though structuring is still visible. We speculate that this more homogenous heating can be caused by heating through ion-neutral collisions, which appear to give rise to a more homogenous distribution of currents and temperature than Spitzer-resistivity only
\citepads{2007ApJ...666..541A,2013ApJ...764...54L}.
We note, however, that systematic biases intrinsic to our inversions might influence the temperature structure on which we base our conclusions.

In this paper we present a global analysis of the heating in the chromosphere of emerging flux regions using multi-line observations at high spatial and temporal resolution.  Future work should focus in detail on individual heating events, such as the bright thread-like  regions of increased temperature that we find in the inversions, as well as a more detailed analysis of the spatial and temporal structure of the temperature in the solar chromosphere.  

Our spectropolarimetric observations in the \CaIIIR\ line have too-low signal to noise ($\sim3\times 10^{-3}$ to derive the horizontal magnetic field in the chromosphere. This highlights the need for imaging-spectropolarimetry in chromospheric lines with a sensitivity at least an order of magnitude better, so that inversion codes can provide estimates of the magnetic field vector at multiple heights in the chromosphere.

\begin{acknowledgements}
The Swedish 1-m Solar Telescope is operated by the Institute for Solar
Physics of Stockholm University in the Spanish Observatorio del Roque
de los Muchachos of the Instituto de Astrof\'{\i}sica de Canarias.
 Computations were performed on resources provided by the Swedish National
 Infrastructure for Computing (SNIC) at the PDC Centre for High Performance Computing (PDC-HPC)
 at the Royal Institute of Technology in Stockholm as well at the High Performance Computing Center North (HPC2N). This research was supported by the CHROMOBS and CHROMATIC grants of the Knut och Alice Wallenberg foundation. JdlCR is supported by grants from the Swedish Research Council (2015-03994), the Swedish National Space Board (128/15). SD and JdlCR are financially supported by the Swedish Civil Contingencies Agency (MSB). MC has received support from the Research Council of Norway.
 \end{acknowledgements}

\bibliographystyle{aa} 
\bibliography{fluxem}

\end{document}